\documentclass[twocolumn,twocolappendix]{aastex63}
\usepackage{amsmath}

\newcommand\kms{km s$^{-1}$}

\defcitealias{kounkel2018a}{Paper I}
\usepackage{listings}

\begin{document}
\title{Supernovae in the Orion: the missing link in the star forming history of the region}

\author[0000-0002-5365-1267]{Marina Kounkel}
\affil{Department of Physics and Astronomy, Western Washington University, 516 High St, Bellingham, WA 98225}
\email{marina.kounkel@wwu.edu}

\begin{abstract}
The Orion Complex is a notable star forming region, that it is fragmented into several different populations that have substantial difference in their phase space. I propose a model that attempts to explain the how the Complex has evolved to this current configuration. In this model, the large scale expansion can be attributable to a supernova that has exploded 6 Myr ago. The remnant of which can be seen as Barnard's loop, as the center of the expansion is consistent with the geometrical center of the HII bubble. This is similar to the HII bubble and the ballistic expansion that is associated with $\lambda$ Ori, a region which has also been a site of an ancient supernova. Assuming that the Orion Complex has originally been forming as one long filament spanning from the bottom of Orion A to $\psi^2$ Ori (or, potentially, as far as $\lambda$ Ori), Barnard's loop supernova could have split the cloud, which lead to the formation of Orion C \& D. Furthermore, the shockwave that has propagated into the filament could have swept along the gas through several pc, which lead to the formation of the singularly most massive cluster in the Solar Neighborhood, the ONC. I also discuss other related nearby events, such as the formation of the Monogem ring, and various runaways that have been ejected from the Orion Complex.
\end{abstract}


\section{Introduction}
The Orion Complex is the closest star forming region that is capable of forming a large number of massive stars. Containing more than 10,000 stars that can be associated with the region \citep{kounkel2019a}, it is a region that has influenced much of our understanding of how young clustered populations form and evolve.

In the recent years, the release of \textit{Gaia} DR2 \citep{gaia-collaboration2018} has significantly improved the precision of the available measurements of the distance and proper motions. Combined with the large number of stars that have been observed with APOGEE to obtain high resolution spectra (and thus, precise radial velocities), this has yielded improved constrains on the 3-dimensional structure and 3-dimensional kinematics of the Orion Complex \citep[hereafter Paper I]{kounkel2018a}. A number of other studies have been conducted with Gaia in Orion, such as, e.g., structure and dynamics of the individual sub-populations \citep{grosschedl2018,getman2019}, expanding the census of the members \citep{chen2019,jerabkova2019a}, and searching for kinematically peculiar stars \citep{mcbride2019,schoettler2020,farias2020},

In addition to Orion A and B molecular clouds - the regions of the current epoch of star formation, the Orion Complex consists of a number of populations that have little molecular gas remaining. Population centered on the $\lambda$ Ori, located at the head of Orion is one such population. Others include Orion C \& D populations are projected on top of one another in the plane of the sky, they have very similar proper motions, however they are separated by $\sim$50 pc in distance, and they have an $\sim$10 \kms difference in the radial velocity of the two populations. Orion C contains the $\sigma$ Ori cluster, and Orion D roughly coincides with the Orion OB1a and Orion OB1b sub-associations, although both extend much further than has been generally assumed prior to the release of \textit{Gaia} DR2 \citetalias{kounkel2018a} 

While all of these populations are kinematically and spatially distinct, they do significantly overlap in the phase space, more alike than they are different, and they do have comparable ages. Although the structure of the Orion is among the most complex compared to all of the star-forming regions in the solar neighborhood  \citep{kounkel2019a}, it is difficult to imagine an scenario in which all of the individual sub-populations don't share a common origin. Nonetheless, a question arises - what has lead to such a peculiar morphology of the Complex? Has it formed in this segmented manner in situ, or has there been an event that fragmented it?

In this paper I propose a model that unifies the star forming history of the Orion Complex that assumes a massive supernova explosion has shaped much of its structure and dynamics. In Section \ref{sec:laori} I review a case study of $\lambda$ Ori supernova. In Section \ref{sec:barnard} I present evidence that of the supernova that is associated with Barnard's loop, for which $\lambda$ Ori supernova can be thought as a scale model. In section \ref{sec:monogem} I highlight the Monogem ring - which is either a supernova remnant associated with a runaway from the Orion Complex, or a signature of bipolar outflow from Barnard's loop supernova. Finally, in Section \ref{sec:disc} I discuss the model of the star forming history of the Complex.

\begin{figure}
\epsscale{1.1}
 \centering
\plotone{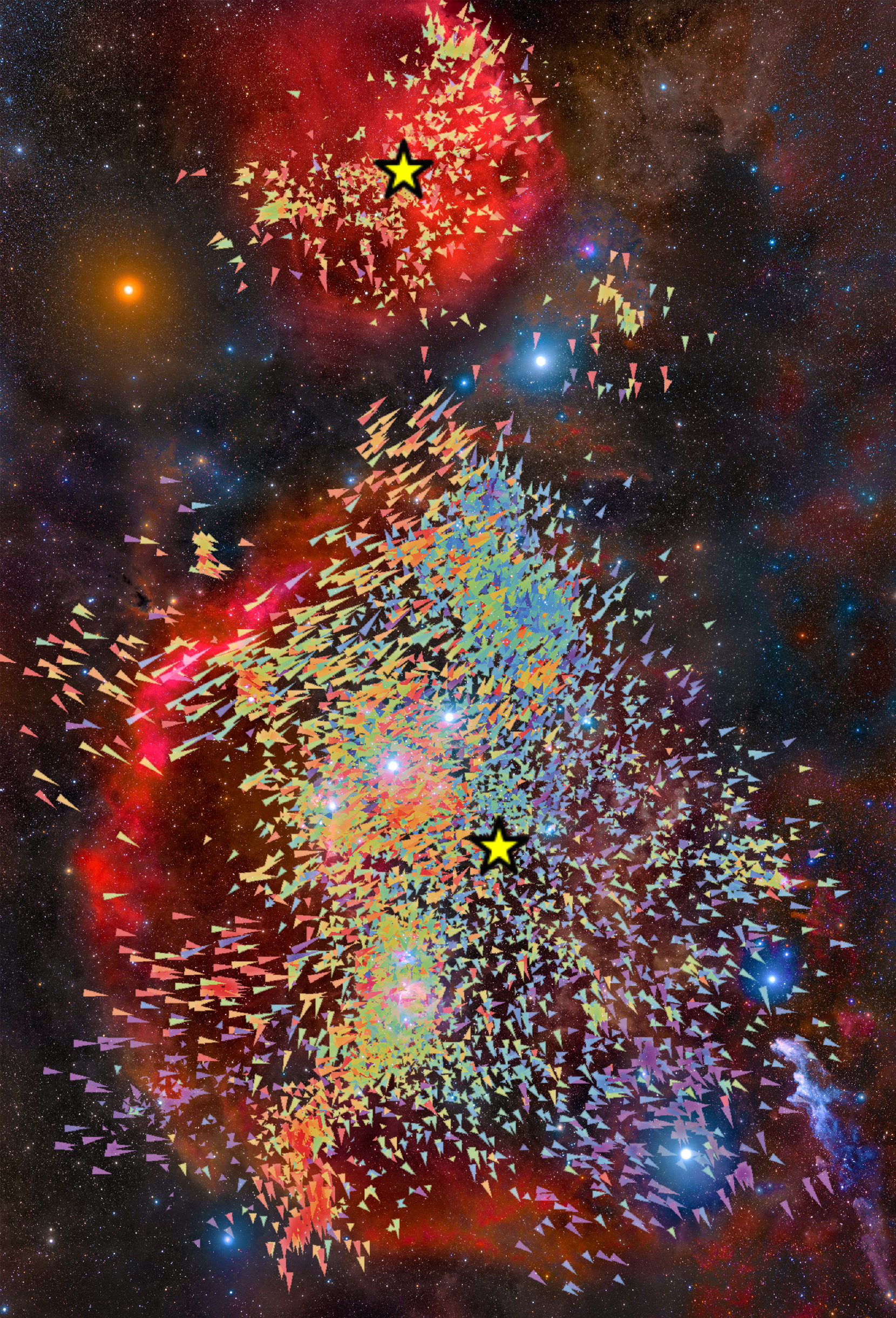}
\plotone{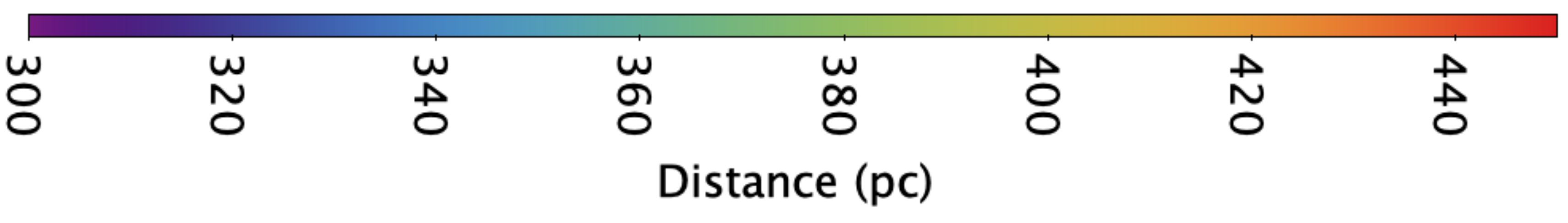}
\caption{Proper motions of the members of the Orion Complex from \citet{kounkel2019a}, color coded by the distance of the stars, from purple at 300 pc to red at 450 pc. Proper motions (pointed from the thicker part of the arrow at the current position, moving towards the thin part) are in the LSR, in the reference frame of the Orion Nebula. Vectors are overlaid on the photograph of Orion, courtesy of Rogelio Bernal Andreo. The two star makers show the approximate locations of the supernova eruptions.
\label{fig:ori}}
\end{figure}

\section{$\lambda$ Ori}\label{sec:laori}

\citetalias{kounkel2018a} has noted that the $\lambda$ Ori cluster has a peculiar kinematic signature. While within the cluster itself appears to be virialized, outside of 1.5$^\circ$, the proper motions of the stars are pointed away from the cluster center. Moreover, the motion is ballistic: the further away they are, the faster they appear to be, up to the speeds of 6 \kms in the rest velocity of the cluster, and they can all be deprojected back towards the cluster center at the age of $\sim$4.8 Myr ago (assuming no acceleration).

$\lambda$ Ori has long since been theorized to be a site of an ancient supernova explosion that went off several Myr ago \citep{dolan1999,dolan2002,mathieu2008}. That supernova has produced an ionized HII bubble. All of the stars associated with this region are firmly enclosed within the bubble (Figure \ref{fig:ori}). Furthermore the stars that are located further towards the edges are younger ($\sim$2 Myr) than those that are found in the cluster center ($\sim$5 Myr).

\citetalias{kounkel2018a} suggested that the ballistic expansion of stars can therefore be thought of a signature of a supernova. The shockwave has rapidly expunched the molecular gas from the cluster, sweeping it along on the radial trajectory. The stars that have subsequently formed from that molecular gas, as it clumped together, maintained the same trajectory. The gravitational feedback of rapidly dispersing gas may have also contributed somewhat in accelerating the stars \citep{zamora-aviles2019}. Although it can be argued what role the shockwave may have had in triggering the subsequent epoch of star formation of the younger population or if those stars have formed regardless of any outside influences \citep{dale2015}, the net result remains the same. While most young clusters tend to expand somewhat, the typical expansion speeds are on the order of $\sim$0.5 \kms\ \citep{kuhn2019}. In simulations, formation of stars that have been triggered from photoionisation due to radiative feedback can impart velocities to a few stars $\sim$1--2 times higher than the natural velocity dispersion of the cloud, but the overall effect is often difficult to measure \citep{dale2015}. Acceleration of stars to speeds an order of magnitude higher than requires an outside influence, most likely attributable to the supernova in some form. And, indeed, the expansion of the shocked gas that is observed around recent supernovae \citep[$\sim$13 \kms,][]{sashida2013} is sufficient to accelerate the gas, and, subsequently, stars, to the necessary speed.

\section{Barnard's loop supernova}\label{sec:barnard}

\begin{figure*}
\begin{interactive}{js}{interactive.zip}
		\gridline{
             \fig{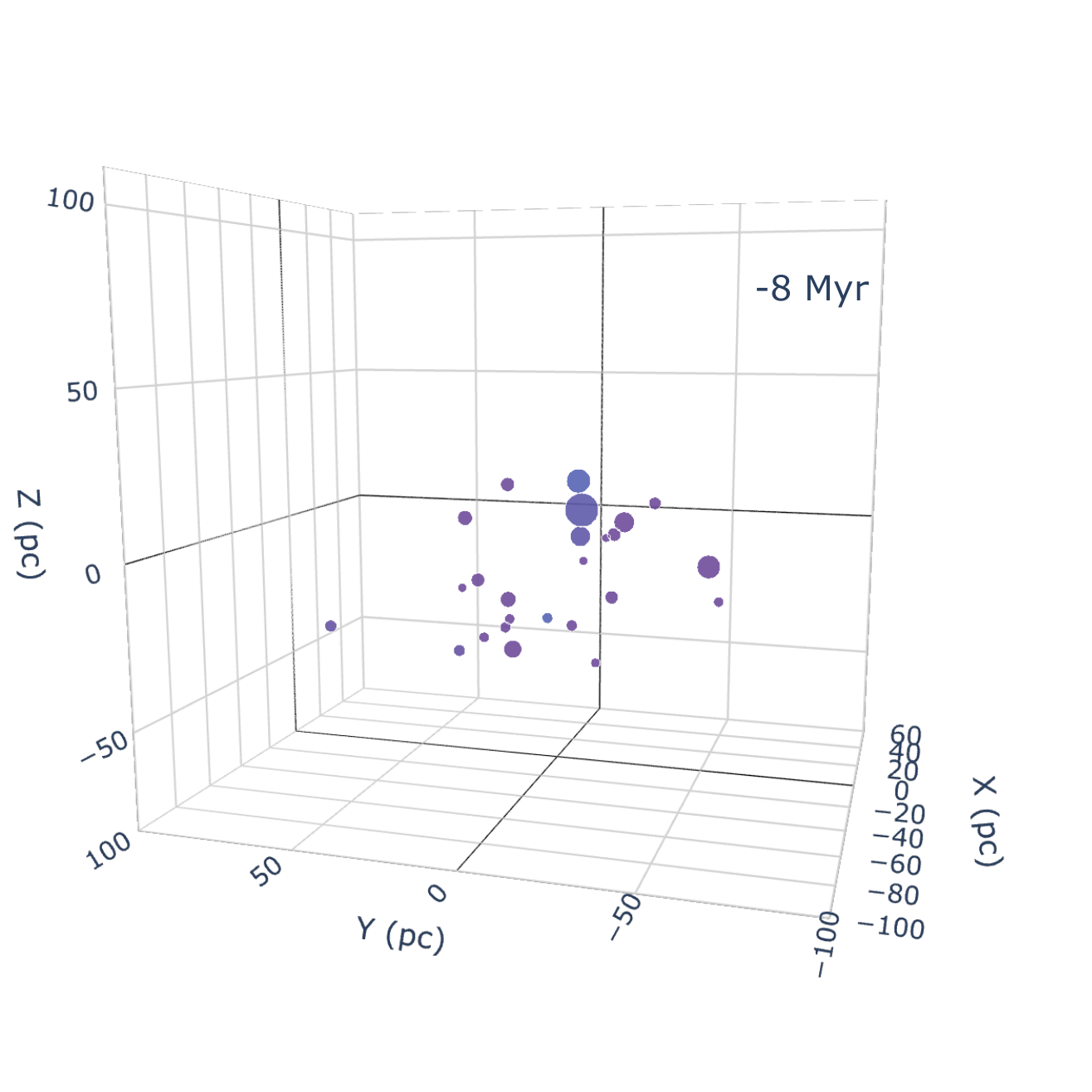}{0.33\textwidth}{}
             \fig{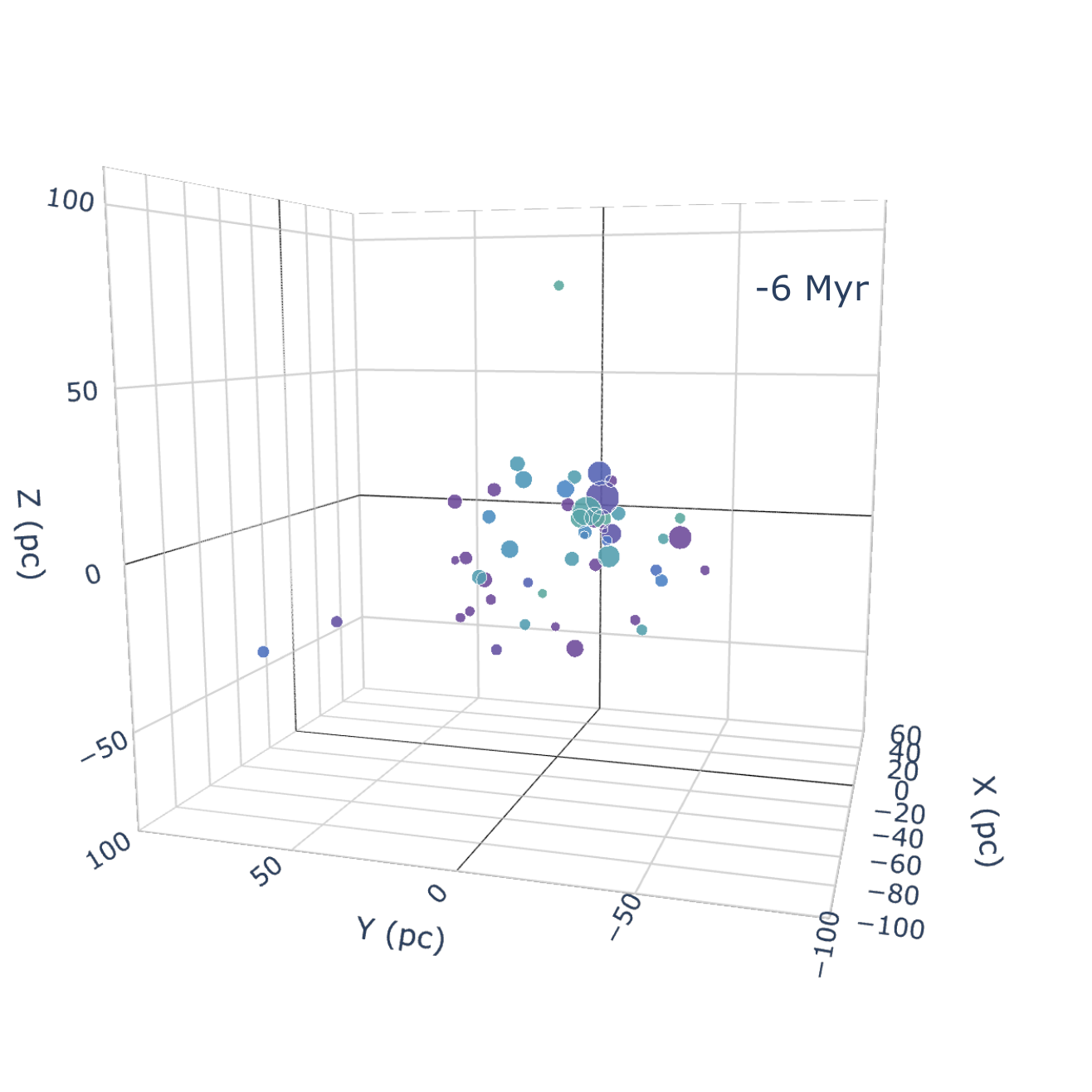}{0.33\textwidth}{}
             \fig{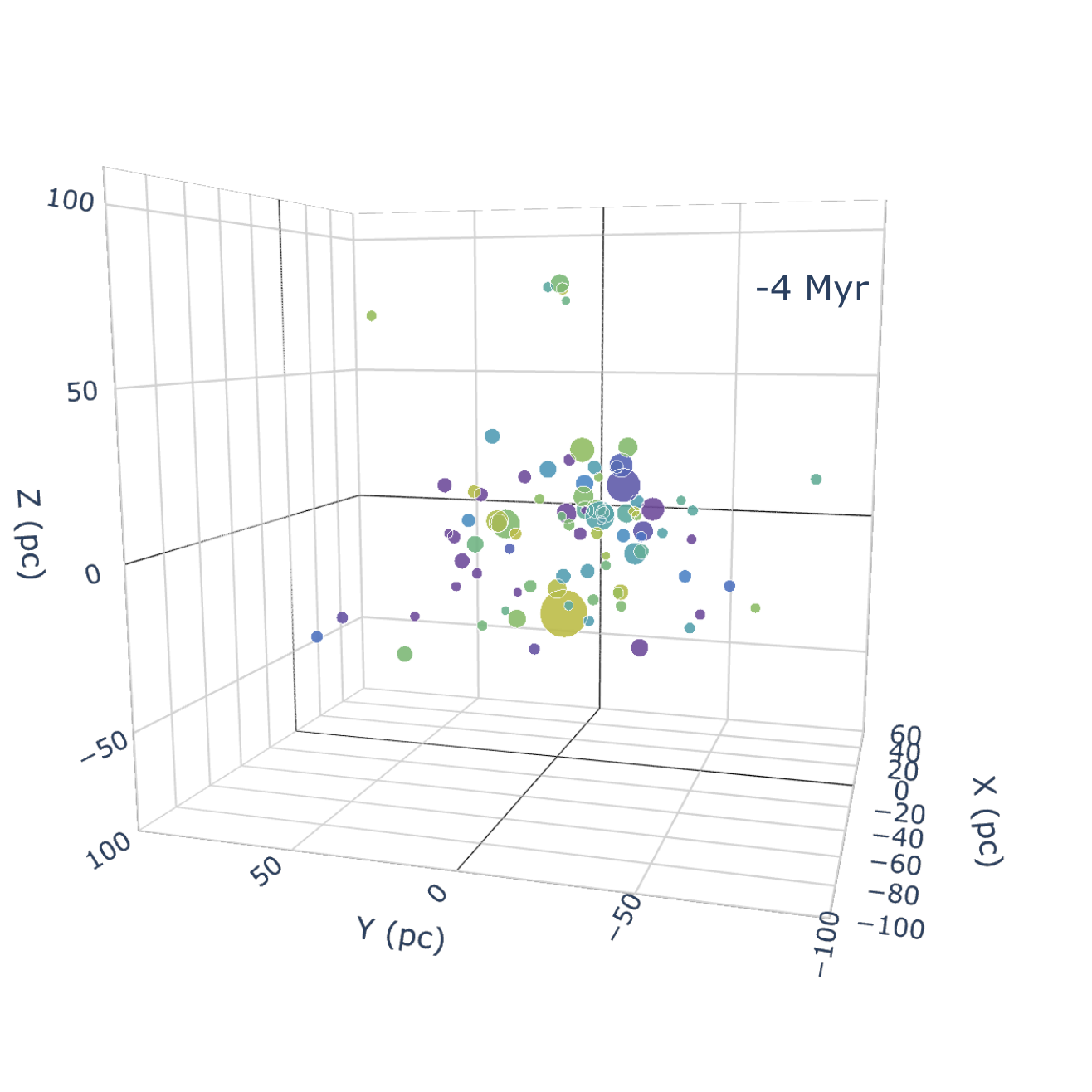}{0.33\textwidth}{}
        }\vspace{-1 cm}
        	\gridline{
             \fig{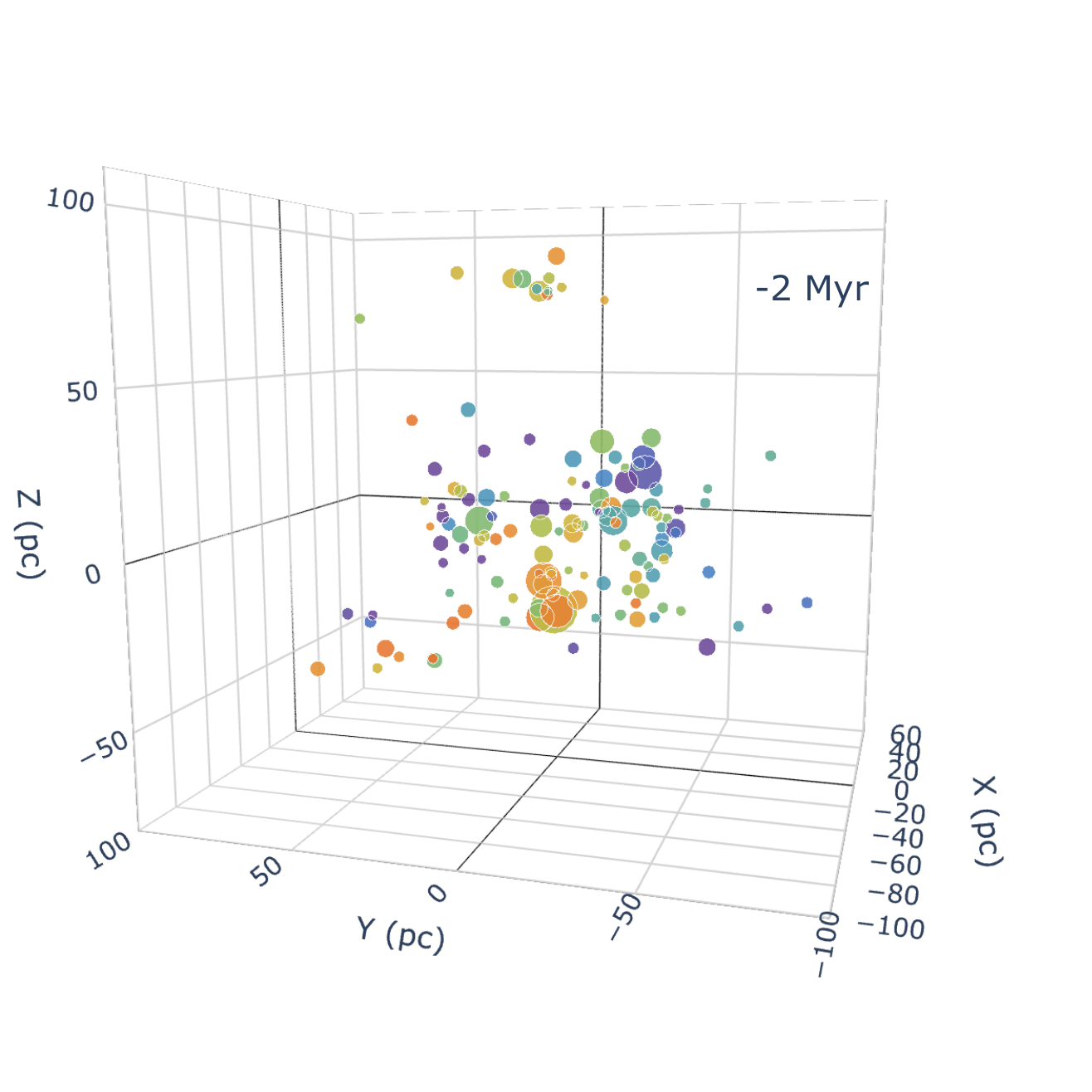}{0.33\textwidth}{}
             \fig{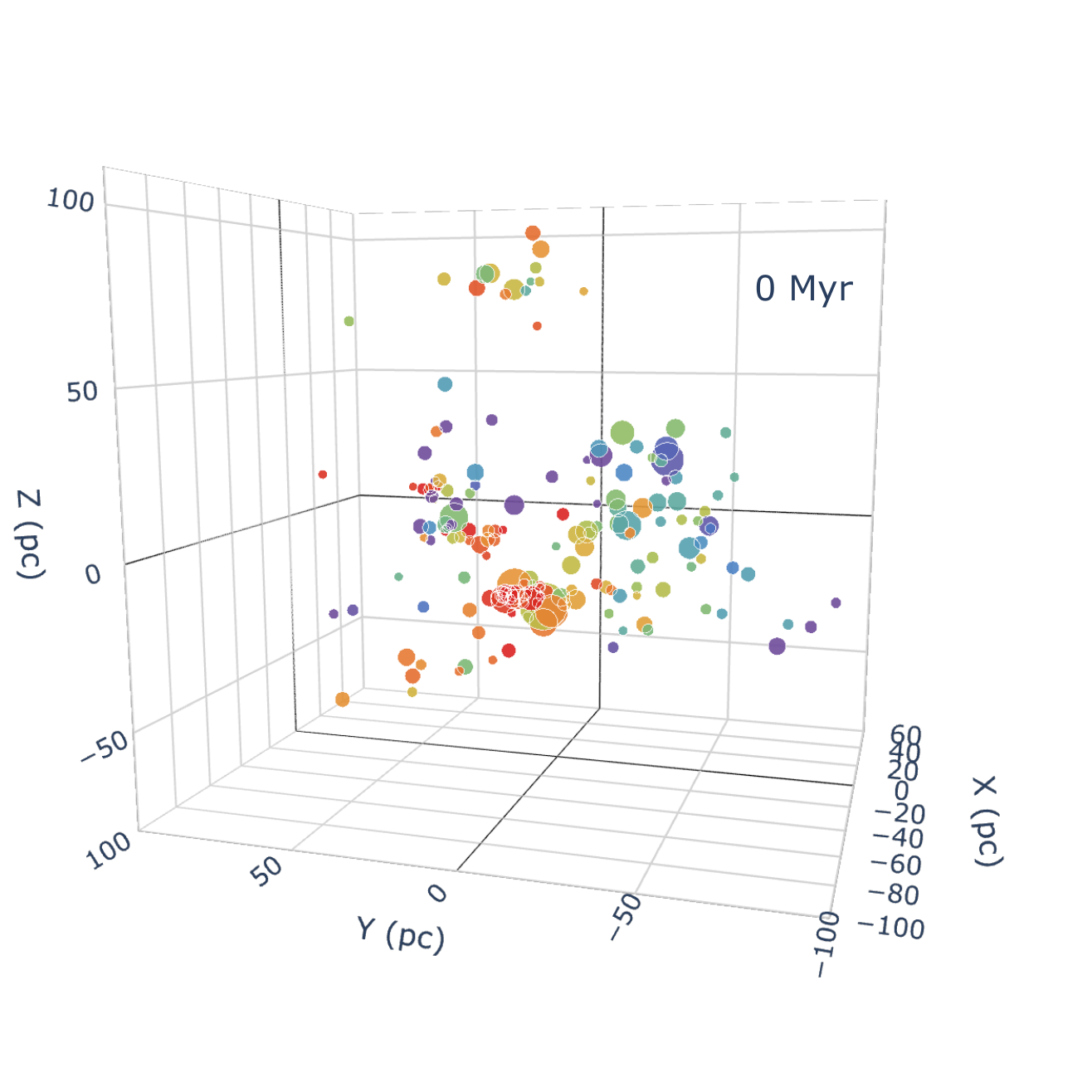}{0.33\textwidth}{}
             \fig{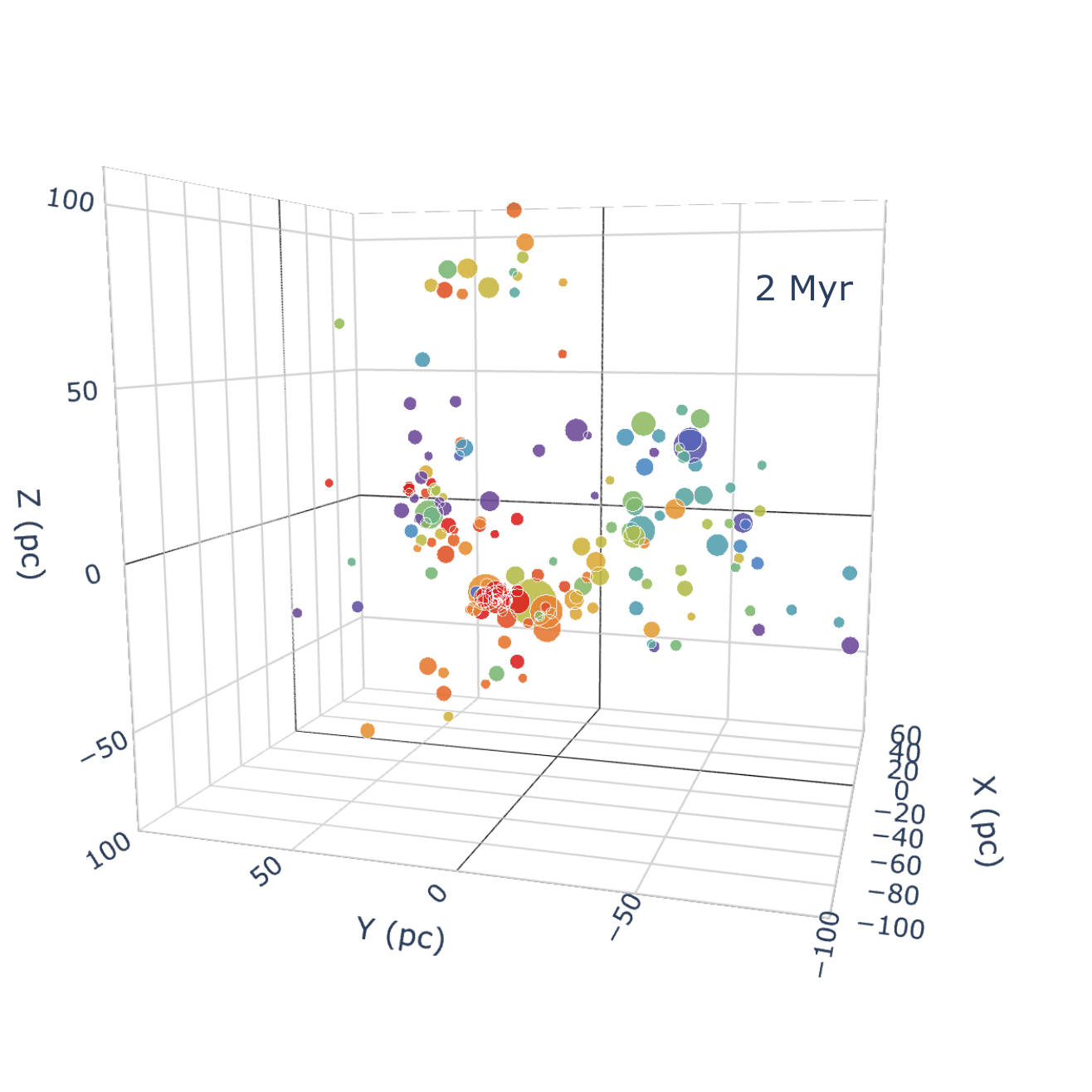}{0.33\textwidth}{}
        }\vspace{-1 cm}
		\gridline{
             \fig{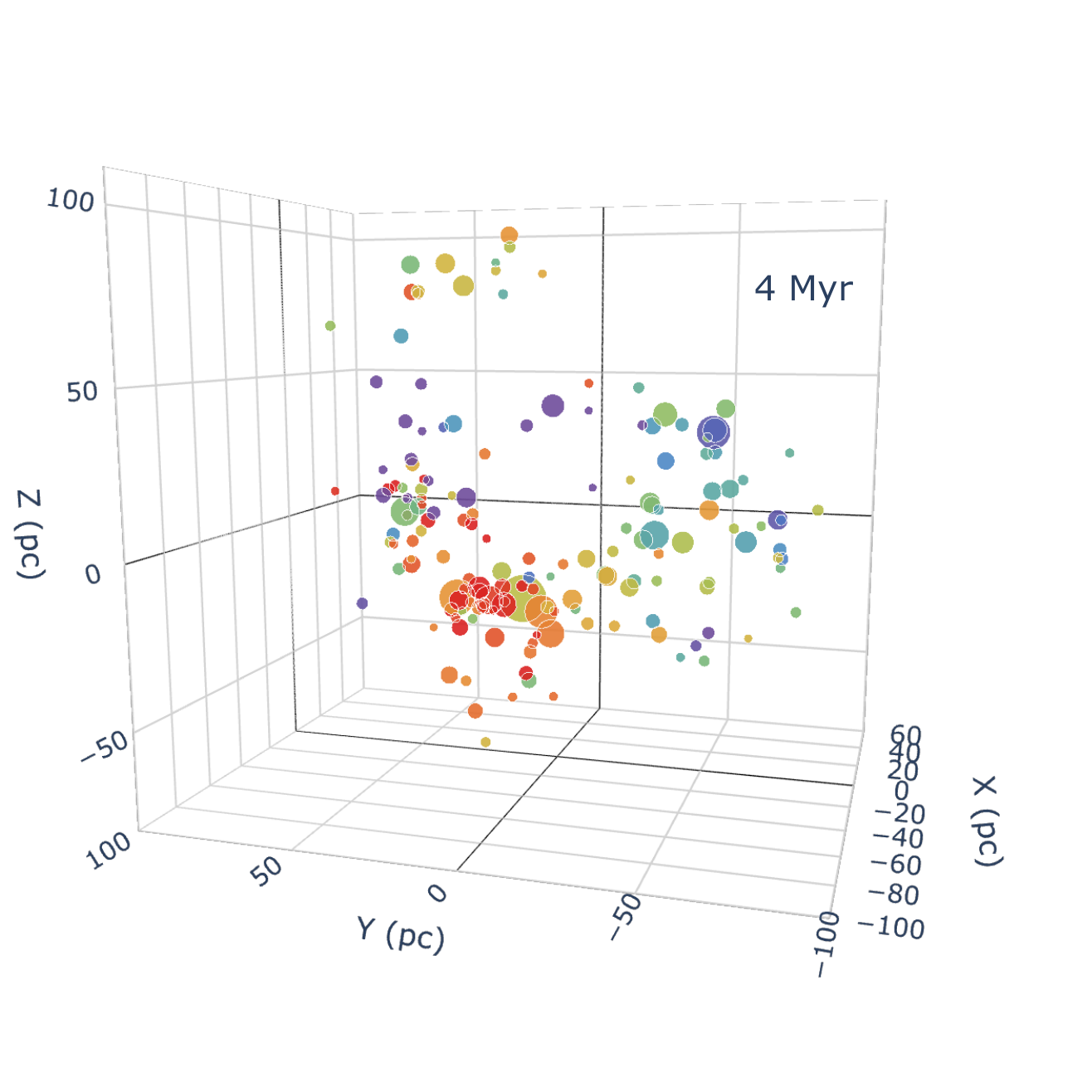}{0.33\textwidth}{}
             \fig{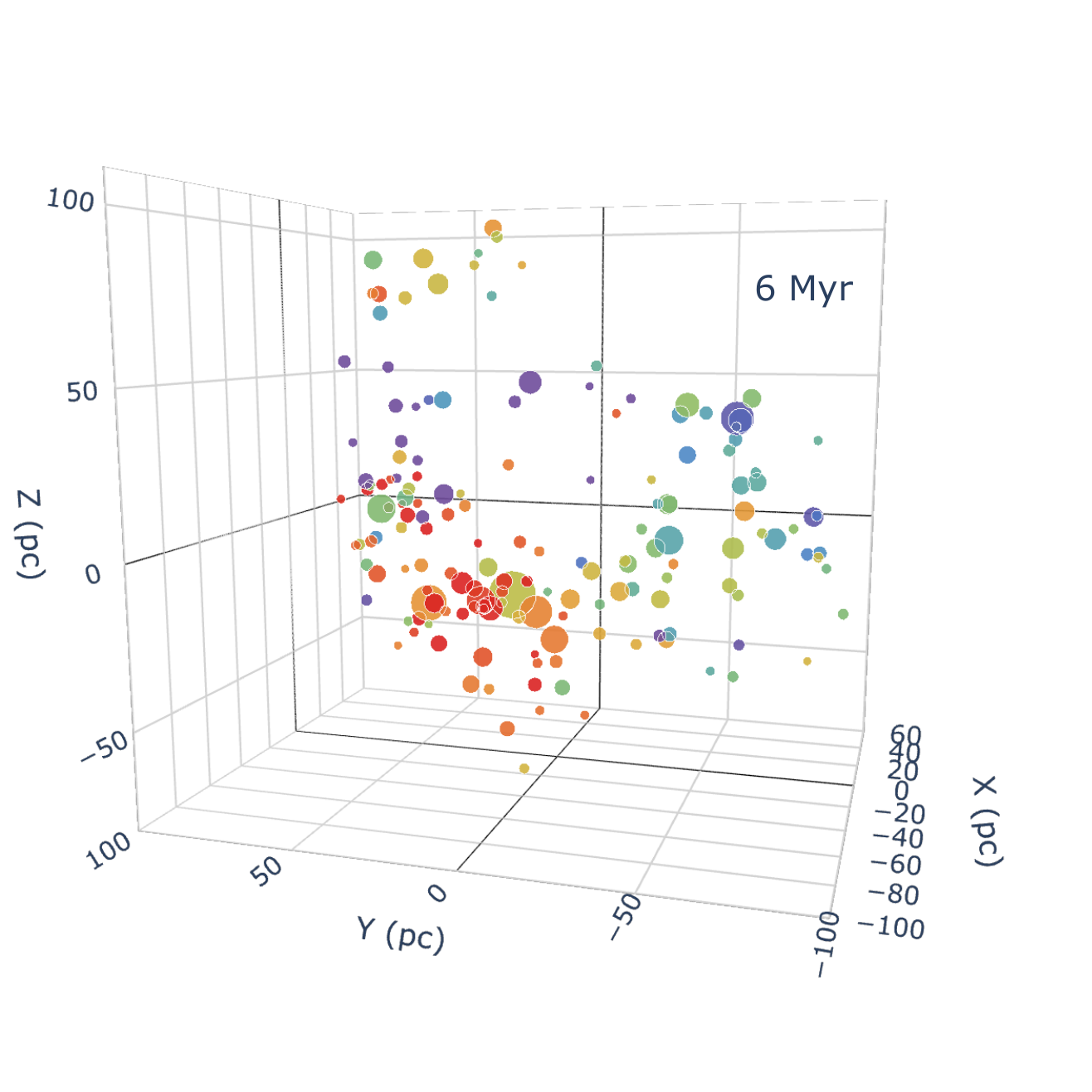}{0.33\textwidth}{}
             \fig{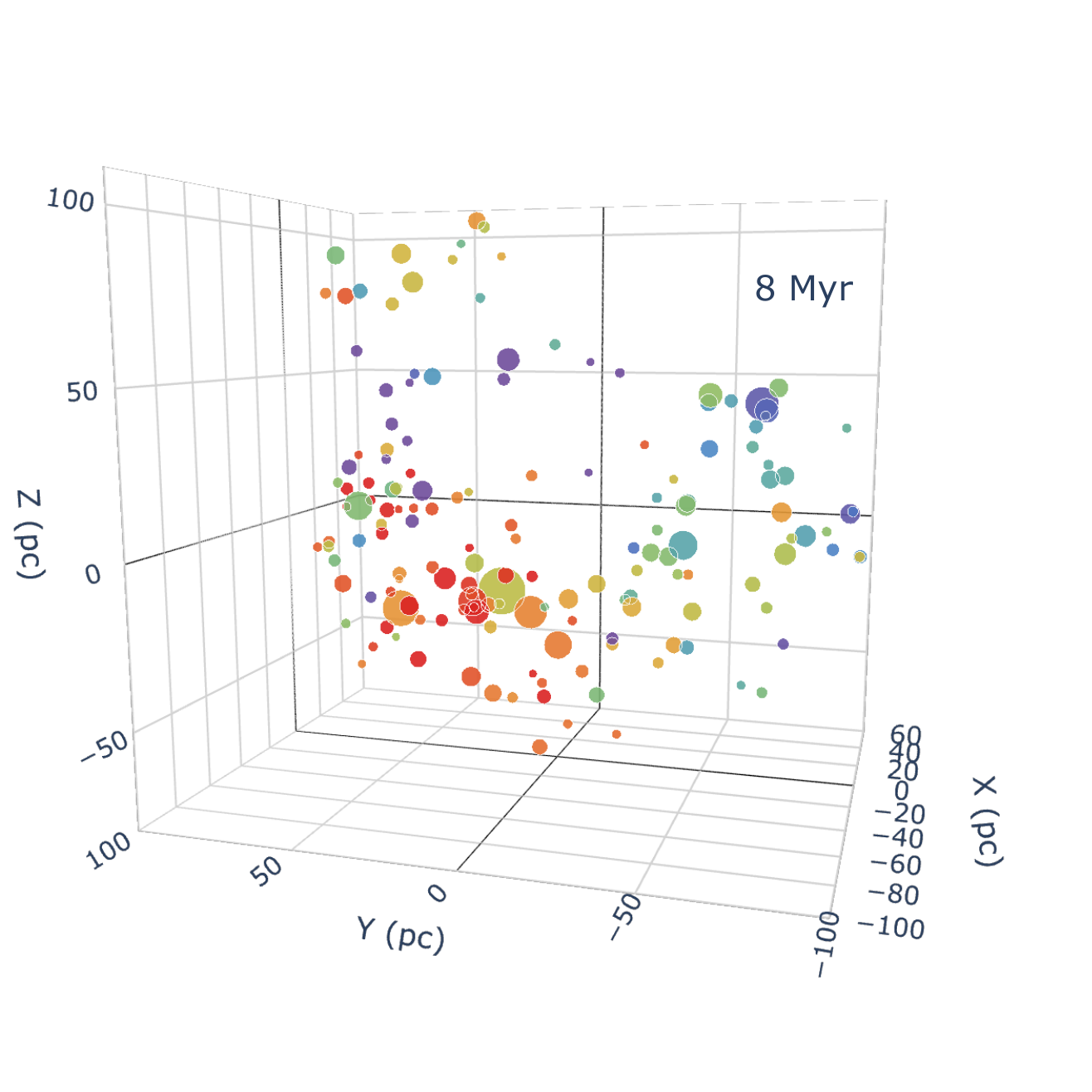}{0.33\textwidth}{}
        }\vspace{-1 cm}
\end{interactive}
\caption{3-dimensional distribution of the groups in Orion from \citetalias{kounkel2018a}. The groups are color coded by the average age of the stars within them with red being the youngest, and the size of the dot corresponds to the number of stars inside each group. The panels show the traceback and the trace forward look over the $-$8 and +8 Myr, with 0 Myr corresponding to the present day. Positions of these groups is linearly evolved through time, assuming the current velocity. The observer is positioned to the right of the image. Interactive version that allows the change in perspective, with the age slider of up to 12 Myr in either direction is available in the online version, temporarily at \url{http://mkounkel.com/mw3d/orionevolution.html}. \label{fig:3d}}
\end{figure*} 

\begin{figure}
\epsscale{1.1}
 \centering
\plotone{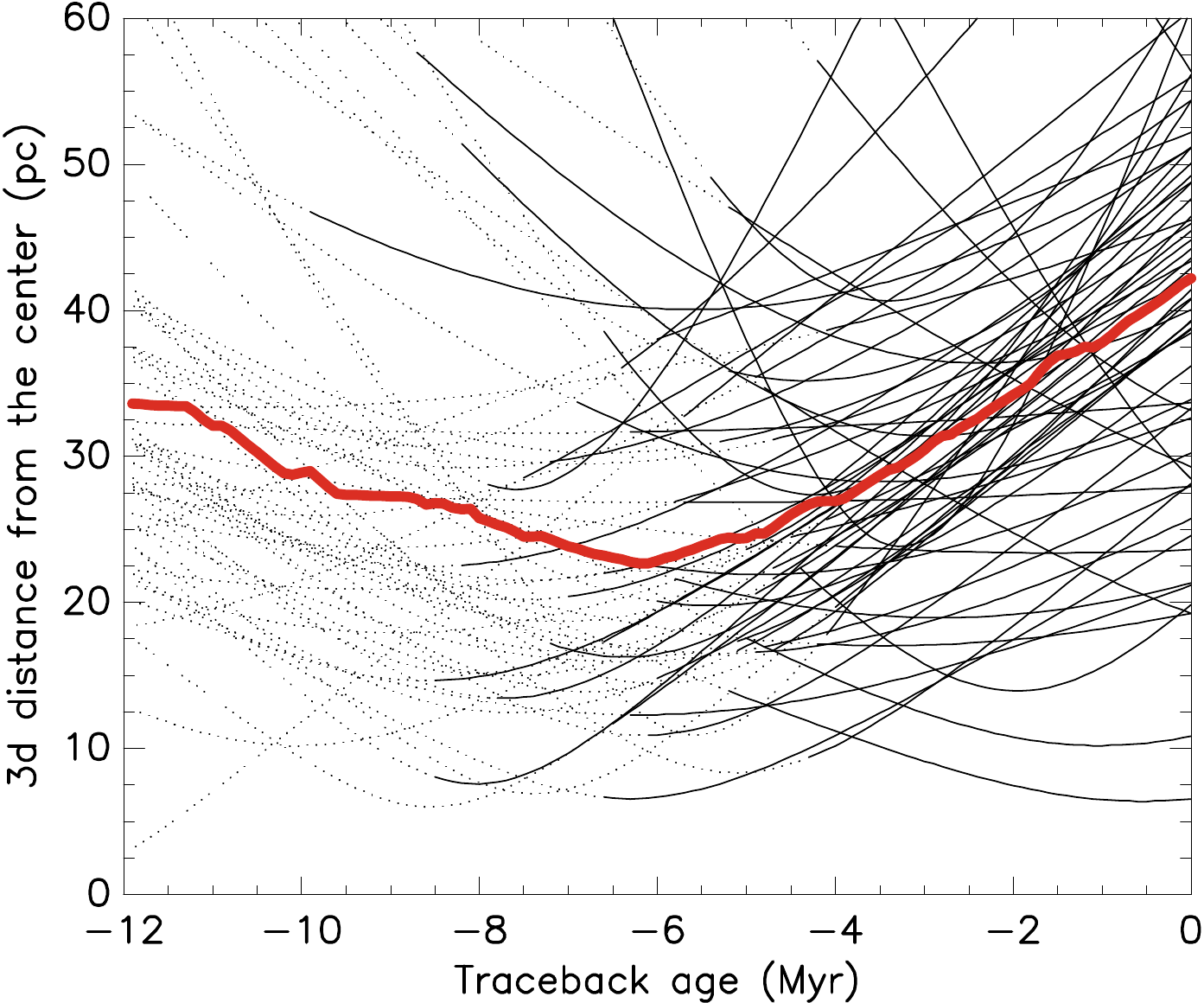}
\caption{Distance of the groups shown in Figure \ref{fig:3d} away from the center of the expansion of Barnard's loop. To remove contamination from the groups unrelated to this expansion (e.g., L1641, $\lambda$ Ori, Orion B, or those that predated the significant formation of stars in the Complex), the groups were limited to be within 40 pc of the center, with the ages between 4 and 10 Myr. Dotted lines show the traceback over the time that predates the formation of the group, solid black lines shows the distance to the center from the time corresponding to the star forming epoch to the present day. Red line shows the median distance of all groups at a given time step.  \label{fig:1d}}
\end{figure}
 
In addition to the HII bubble associated with $\lambda$ Ori, there is another very notable bubble: Barnard's loop. It is unknown what the origins of it are, but it has been theorized that it may have been a byproduct of a supernova \citep{madsen2006,ochsendorf2015}, or that it has been driven by the radiation pressure of the OB stars \citep{odell1967,odell2011}. Similarly, there has been some debate whether Barnard's loop is just a part of the Orion-Eridanus superbubble \citep{wilson2005}, or if an independent entity \citep{ochsendorf2015}.

Taken as a whole, much of the Orion Complex is perfectly encircled by Barnard's loop: from Rigel to $\psi^2$ Ori, from the top of Orion B to the bottom of Orion A, not dissimilar to the $\lambda$ Ori bubble. As such, it is highly likely that they are associated with one another. The geometrical center of the bubble lies approximately a degree away to the southwest of $\eta$ Ori. 

In \citetalias{kounkel2018a} we noted that Orion D appears to be expanding, attributing it as a natural evolution due to the age of the population. But, considering that young populations of similar mass can survive as comoving groups for several hundreds of Myr \citep{kounkel2019a}, a 8 Myr population tearing itself apart in such a manner appears to be unusual. Upon closer examination I found that when all the proper motions for the entire Orion Complex are placed in the common reference frame, the center of the expansion appears to correspond to the geometrical center of Barnard's loop (Figure \ref{fig:3d}, and that many of the stars that are moving radially away from it can be traced back to this origin $\sim$6 Myr ago.

Currently, Orion C and Orion D currently exist as separate entities, located at two very different distances, of 412 pc and 350 pc respectively, on average. They are colocated along the recenttly discovered bubble of dust \citep{rezaei-kh.2020}. However, as Orion C has a radial velocity of $\sim$13 \kms, and Orion D has a radial velocity of $\sim$4 \kms\ (in the local standard of rest (LSR) reference frame), they are receding away from each other, and, in the past, they would have been much closer together. They would have been co-located $\sim$6-7 Myr ago at the distance of $\sim$320--330 pc, not dissimilar to the distance the Orion Nebula Cluster (ONC) would have had at the time (projecting back the current distance of 389 pc with the characteristic RV of 10 \kms), ignoring the relative motion of the Sun to the LSR. Similarly, prior to the expansion, their average rest frame radial velocity would have been similar to that of the ONC as well ($\sim8$ \kms). The timescale of the expansion along the line of sight is well matched to the expansion in the plane of the sky.

\citet{grosschedl2018} have noted that Orion A molecular cloud has a peculiar shape, that the ONC, the "head" of the cloud is tilted relative to it, and that as if pushed by some force perpendicular to the filament. They proposed that it could be attributable to either cloud-cloud collision, or due to stellar radiative and supernova feedback. \citet{getman2019} have also suggested that the compression shock from a feedback from an OB stars from Orion D could be responsible for the dynamical evolution of the Head, and that this compression then assisted in the global gravitational collapse of the cloud.

The direction of the compression of Orion A (based on the 3d map, Figure \ref{fig:3d}) is consistent with originating from the center of the bubble.

In \citetalias{kounkel2018a}, we performed hierarchical clustering of the Orion Complex to separate it into $\sim$200 groups that trace the full extent of the Complex, each one representative of the position and velocity of the stars in a particular sub-region. These groups can act as tracers of the dynamical evolution of the Complex. I used the groups that have complete phase space information, i.e., radial velocities either from APOGEE or from Gaia are available for some stars in the group. \citetalias{kounkel2018a} measured the average age of the stars for each group - $Age_{HR}$ was used, when available (as it has been extinction corrected), otherwise, $Age_{CMD}$ was used. In Figure \ref{fig:3d} I show a 3-dimensional traceback model of Orion, where different groups are at a particular time, ranging from 12 Myr in the past, to 12 Myr in the future (in the interactive version of the plot). Different groups are added only added at the time that corresponds to their age and are excluded from the previous timestamps.

This traceback is quite simplistic. It does not account for the self-gravity of the Complex, either stars or gas. No two groups able to interact, even when they are both components of the same cluster (i.e, although the subgroups composing the ONC appear to fly apart in the traceback going far enough into the future, it is unlikely to happen in reality). Furthermore, no new stars are able to form. Nonetheless, it does show the general patterns of motion, that the expansion of the Complex all indeed appears to originate from the same point, and that it is largely spherical, with the dominant plane for the stellar distribution.

Figure \ref{fig:1d} further demonstrates it, reducing the 3-dimensional motion of the individual groups to 1d distance to the apparent center of the expansion. Although they do not necessarily all converge to zero (due to uncertainty in parallax, uncertainty in correction for the average position and velocity, physical sizes of the groups themselves, velocity dispersion within the group, self gravity of the Complex affecting the trajectory, as well as the initial geometry of the cloud), the minimum size of the expanding populations indeed occurs $\sim$6 Myr ago. Furthermore, the majority of star formation has occurred after this time.

Putting all of these pieces together, it becomes evident that the radial expansion of the Orion Complex from the geometrical center of Barnard's loop and the formation of the HII bubble itself is likely to be attributable to the same event. Moreover, there are many similar hallmarks in the velocity structure and the HII bubble compared to $\lambda$ Ori. From these dynamical I propose that a supernova that erupted $\sim$6 Myr is the most likely cause.

As is the case with $\lambda$ Ori, supernova is the primary event that is capable of accelerating the gas and stars to the appropriate speeds and produce the structure of the current size, as other forms of stellar feedback do not have sufficient force. Although a number of other HII bubbles are present in the Orion Complex that are caused by photoionization and winds from OB stars, they tend to be only a few pc in diameter and have a clear driving source. A lack of a suitably bright and massive star at the epicenter of the expansion despite a clear spherical geometry further suggests that while such a driving source most likely have existed in a past, it has since died off.

Recently, \citet{grossschedl2020} have also independently identified a episode of major feedback event in the Orion Complex dating back 6--7 Myr ago, which they refer to as Orion-6 event, using different tracers. They suggest that the most likely origin of this event is the Orion X region, which is an overdensity of stars in Orion D located somewhat south of $\eta$ Ori. Although, currently, Orion X is located in one of the expanding shells, it's line of sight is indeed consistent with the center of the bubble, and 6 Myr ago it would have been much closer in the 3d space as well. Better determination of stellar ages is needed to confirm if it could have formed the progenitor, or if it has been one of the first regions formation of which would have been assisted by the shockwave.

However, rather than a single event, \citet{grossschedl2020} suggest that it might have been a result of multiple triggers that might have continued over time, to explain the difference in momentum of the different clouds. While self-gravity of the Orion Complex may explain some of these differences, it may be difficult to discriminate between a single explosive event and a group of neighboring events that occurred close in time. Future simulations, such as some of the ones noted in \citet{grossschedl2020}, would be able to more definitively address this point.

\section{Monogem Ring}\label{sec:monogem}

\begin{figure}
\epsscale{1.1}
 \centering
\plotone{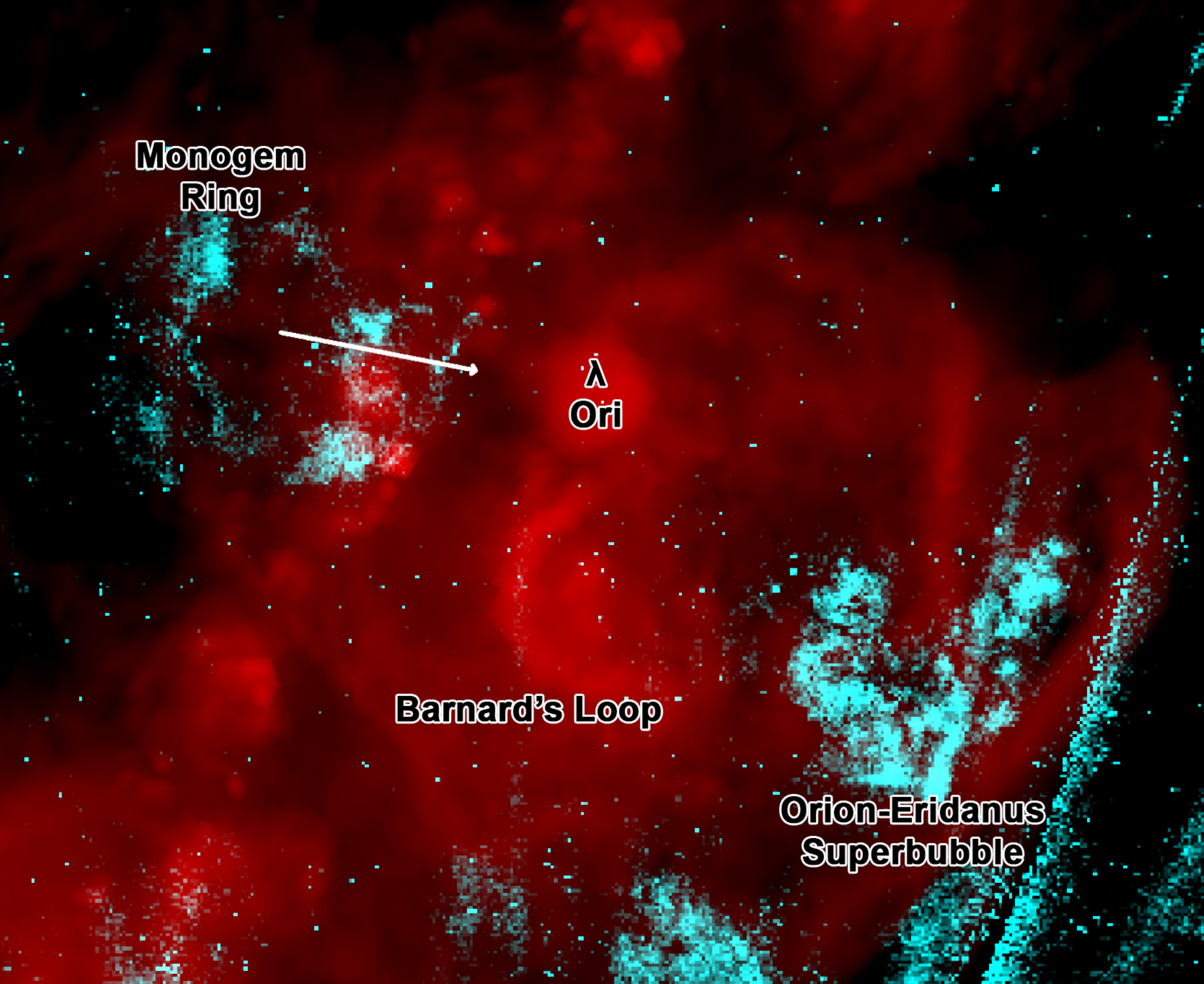}
\caption{ROSAT X-ray emission map \citep[cyan,][]{snowden1997}, superimposed on the H$\alpha$ map \citep[red,][]{haffner2003}. The arrow shows the traceback projection of PSR B0656+14 over the last $\sim$1 Myr, assuming its current (LSR) proper motions \citep{golden2005}.
\label{fig:xray}}
\end{figure}

In examining the ROSAT X-ray maps of the region \citep{snowden1997}, two features are apparent (Figure \ref{fig:xray}). One, located to the west of Orion, is the emission associated with the Orion--Eridanus Superbubble, and it is thought to be related to the winds or the supernovae from the Orion Complex \citep{ochsendorf2015}. The bubble is found at the distance similar to the Complex, and low below the galactic plane that no other population can be found that could serve as a likely progenitor. Indeed, even more complex clustering analyses akin to \citet{kounkel2019a} do not reveal any overdensities in the phase space inside the superbubble.

Another X-ray feature is located to the east of Orion, and it is the Monogem ring. It is commonly thought of as a supernova remnant \citep{plucinsky1996,knies2018} with an estimated age of 0.068 Myr. It has a pulsar near its center PSR B0656+14 with a spin down age comparable to this estimate \citep{thorsett2003}. Both Monogem and PSR B0656+14 are thought to be located at the distance of $\sim$300 pc \citep{golden2005}.

Similarly to the Orion--Eridanus Superbubble, due to its height above the Galactic plane, no young stellar population in vicinity that could serve as progenitors can be associated with Monogem. Although some stars have been proposed to form such a population \citep{knies2018}, with the revised \textit{Gaia} astrometry to them compared to Hipparcos, no coherence in their phase space is apparent.

Nonetheless, if we assume that the proper motions of the pulsar measured with VLBA \citep{golden2005} are representative of what the star that produced it had originally, then it may have originated from the Orion Complex as a runaway. Converting the proper motions to the LSR reference frame, the most likely origin is the $\lambda$ Ori Cluster, $\sim$1.7 Myr ago. Thus, if Monogem is a supernova remnant, it is also related to Orion.

PSR B0656+14 is the second pulsar that could be traced back to Orion. Geminga is also a neutron star that likely have formed from a dynamically ejected runaway, originating from 25 Ori cluster $\sim$1.3 Myr ago \citep{faherty2007}.

The relative configuration of Monogem and Orion-Eridanus superbubble, with the Orion Complex positioned equidistantly in between of them, $\sim$200 pc apart, does appear to be somewhat peculiar. The two are located at similar distances from Earth as well, and, have a somewhat similar morphology and size on the sky. They are evocative of a bipolar outflow that may have been associated with Barnard's loop supernova. Further investigation would be needed in the future to explore this possibility. In particular, the increased sensitivity, as well as spectral and spatial  resolution of eROSITA may be of benefit in establishing relationship between these two features.

\section{Discussion}\label{sec:disc}

\begin{figure*}
		\gridline{
             \fig{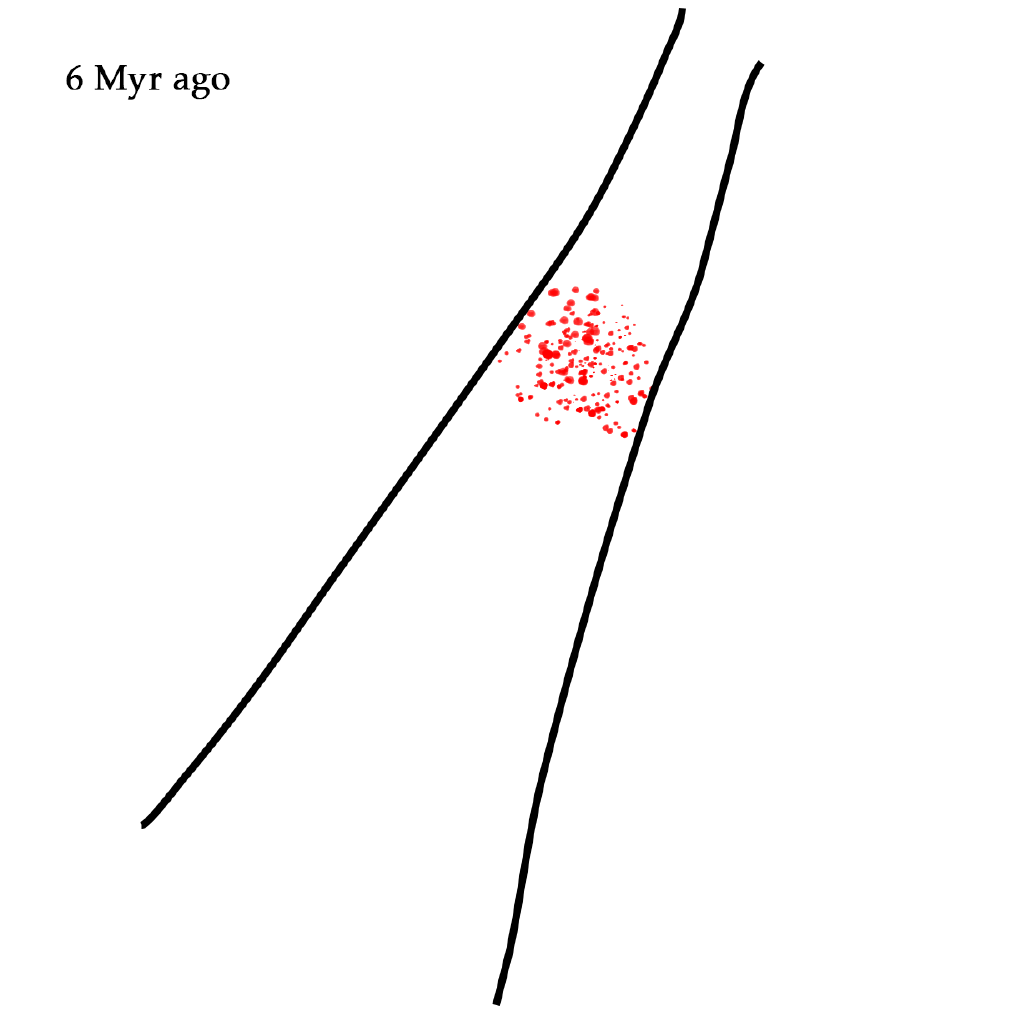}{0.33\textwidth}{}
             \fig{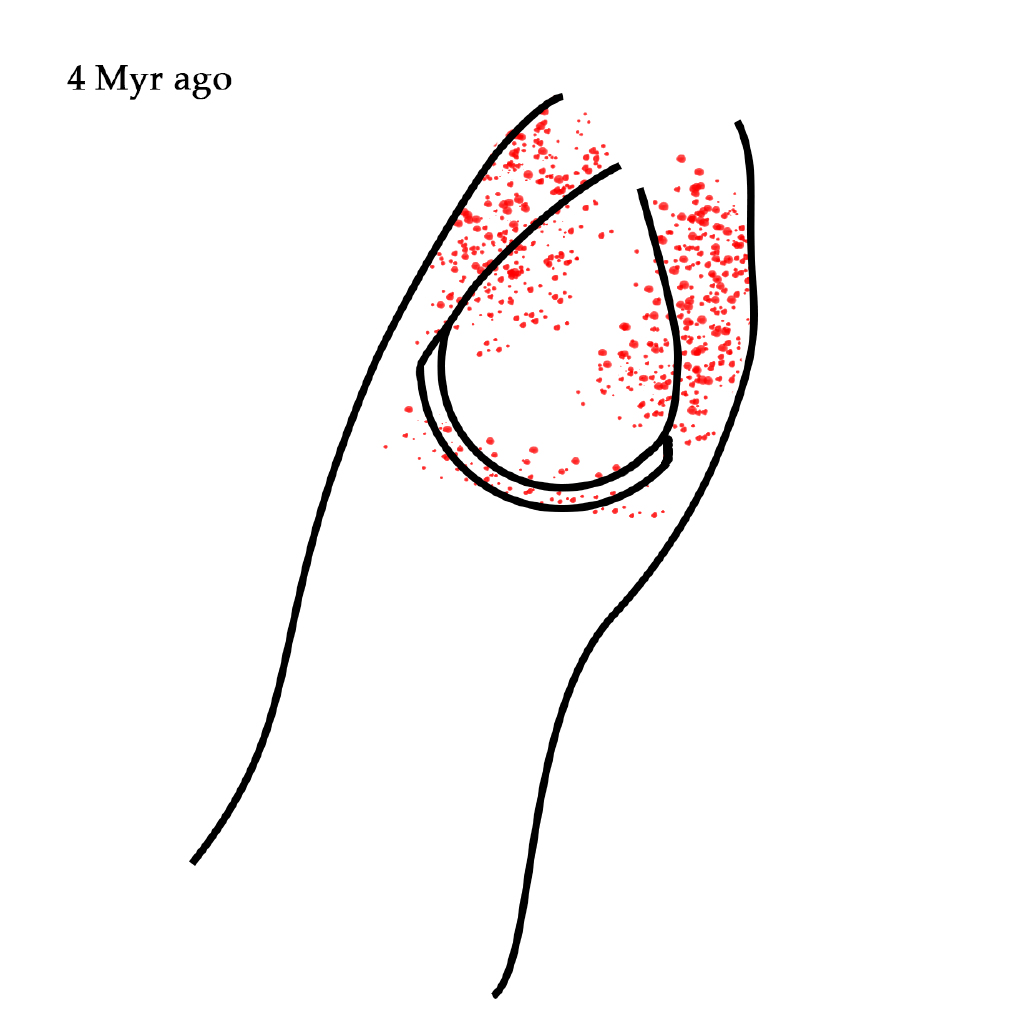}{0.33\textwidth}{}
             \fig{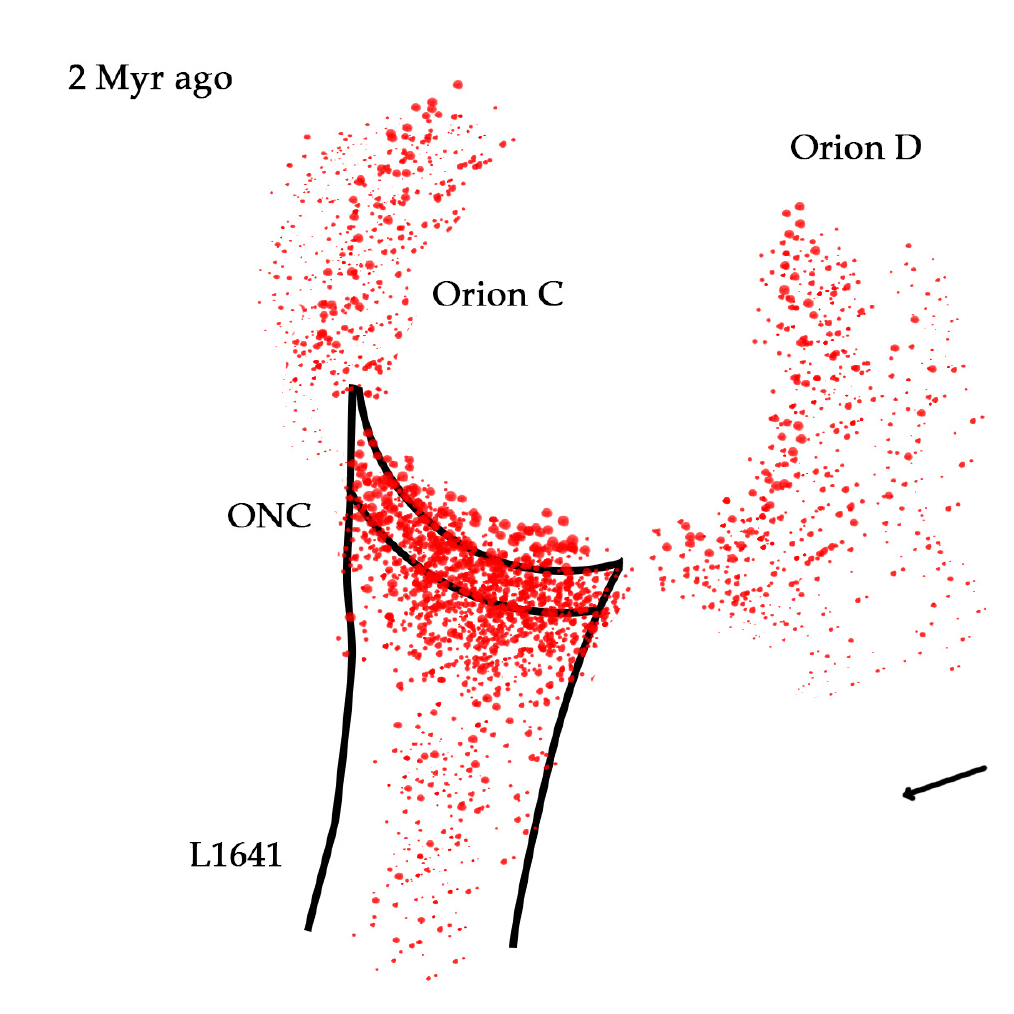}{0.33\textwidth}{}}
\caption{A conceptual model of the formation of the Orion Complex, excluding Orion B and $\lambda$ Ori. Black contours represent the molecular gas, red dots are the stars. The observer is located to the right of the image, direction indicated by the arrow. Left: early epoch of star formation in the filament. Middle: post supernova eruption, the filament begins to be split and compressed. Star formation in Orion C \& D continues as they separate. Beginning of the compression of the gas in the direction of the filament, seeding the formation of the ONC. Right: The molecular gas in Orion C \& D has been largely dissipated, as the stars inside them continue spherical expansion. The ONC becomes massive. L1641 becomes dense enough to begin forming stars. \label{fig:model}}
\end{figure*}

It is notable that the stellar density distribution of the Orion Complex appears to be continuous from the bottom of Orion A, up to $\psi^2$ Ori (possibly up to $\lambda$ Ori, after a small gap), as though forming one long filament. This filament is made less apparent by the stars that are expanding away from the Complex, near Rigel and near L1616, although they would have originated from the filament also. Furthermore, there are a number of outlying clouds and populations -- some fairly massive (namely Orion B), some significantly less so (see Figure \ref{fig:ori}) that are infalling towards the Complex. Although they are excellent examples of the gravity at work, they act as deterrents in visualizing a simplistic model of the Complex, requiring a conceptual separation of these regions, as their formation and their kinematics are driven by different mechanisms compared to the main filament.

Because Orion A and B molecular clouds have been a cornerstone in our understanding of star formation, it is easy to think of them as discrete and complete units, the entire clouds coalescing at the same time, that what we see of them now is all there ever was, and that while other clouds would have existed in the vicinity, they would have always been separate entities. However, recent studies of the solar neighborhood have found a number of stellar strings -- large structures composed of comoving stars extending for several hundred pc in length and only a few pc in width, resembling filamentary molecular clouds from which these stars likely to have formed and retained their morphology long after the gas is dispersed \citep{kounkel2019a}. Moreover, these strings appear to be a dominant form of star formation, accounting for most of the stars that are found in comoving groups up to an age of $\sim100$ Myr. With this in mind it becomes possible to imagine the entire Orion Complex as one string formed from the same molecular filament (or, perhaps, a narrow sheet of gas). That while Orion A is the southern part of this filament that still exists in the molecular gas form, and it had recently has reached densities large enough to start forming stars, the northern part of the same filament has became Orion C \& D and has dispersed its gas. On the other hand, Orion B is quite distinct from the rest, not part of the same filament, but rather, it's in the process of infalling towards it.

The shockwave of the erupting supernova would have been able to sweep the molecular gas and segment the filament. The northern part has already started actively forming stars for $>$2 Myr at that point. Most likely, the deceased star has been among those newly formed, although it could have been a left-over from the previous generation of star formation in the region \citep[e.g.,][]{jerabkova2019a}.

Whether it is due to asymmetries in the distribution of gas, the location of star relative to the filament, or both, the gas has been propelled primarily in 3 directions, producing the current geometry of the Orion Complex that has analogues in the simulations of supernovae in giant molecular clouds \citep[e.g.,][]{smith2020}. Two shockwaves were propelled towards the front and towards the back of the filament, relatively rapidly consuming already partially depleted gas as Orion C \& D begin to take form as separate entities. The gravitational feedback \citep{zamora-aviles2019} may have also assisted in this process. Meanwhile, the third shockfront has been pushed back into the southern part of the filament that was at the time just beginning to collapse (Figure \ref{fig:model}). Although its original density likely been comparable in what is found in L1641, the shockwave could have accumulated the gas from several pc into a single concentrated region. The line-of-sight separation between the ONC and the proposed eruption site is $\sim$10 pc. Such a rapid piling of gas could have lead to the formation of the singularly most massive young cluster in the Solar Neighborhood.

This could also explain the uneven distribution of ages in the cluster \citep[i.e, while older stars are found throughout the head of Orion A, the younger ones are concentrated at the center of the cluster][]{beccari2017}. Although it can be argued whether these generations of stars correspond to the discrete events \citep{jerabkova2019}, or that the distribution of ages is more continuous \citep{da-rio2010, olney2020}, as the cluster continued to sweep through the filament, it would have had access to more gas to support several generations of stars, in a ``conveyor belt'' manner \citep{krumholz2020}. And, although the initial sweep of gas would have allowed star formation along the entire width of the filament, subsequently the self-gravity of the forming cluster would have become more important.

Eventually, as the density of the accumulated gas would have been sufficiently high, and the pressure behind the shockwave decreased as it expanded out further, the momentum of the ONC traveling through the filament was able to decrease, allowing the front to pass through.

The model does not explain the formation of Orion B molecular cloud or other less massive infalling clumps of gas. Orion B is the only one of these clouds through which the shockwave (as traced by Barnard's loop) would have transversed. But due to its distance from the epicenter of the explosion, the effect would have been significantly weaker and difficult to conclusively observe due to feedback from $\sigma$ Ori on the cloud acting from a similar direction.

I propose the following timeline for the events (Figure \ref{fig:model}:
\begin{itemize}
\item \textit{$>$8 Myr ago: }first stars of the Orion Complex form
\item \textit{$\sim$6 Myr ago: }Barnard Loop supernova explodes. Beginning of the formation of the $\lambda$ Ori cluster at a similar time north of the Complex.
\item \textit{$\sim$4 Myr ago: }shockwave expands, splitting the cloud. ONC is starting to form at one of shock-fronts as the gas as the filament begins to pile up. Orion C \& D continue to form along the opposing shockfronts. At a similar time, $\lambda$ Ori supernova explodes.
\item \textit{$<$2 Myr ago: }The split between Orion C and D becomes more pronounced. Gas on the periphery of the Complex continues to infall and collapse onto it.
\end{itemize}

Unfortunately, no remnant is known to be associated with Barnard's loop (as with $\lambda$ Ori). Most of the known neutron stars, including Geminga and PSR B0656+14 are less than 1 Myr old. And, as even at those ages they tend to be faint, only a few have reliable astrometry. If a remnant is a black hole (which may be likely, given the duration of time between the formation of the progenitors and their death), then they are even more difficult to detect. Dedicated X-ray surveys of the likely site of the explosion may help to shed the light on a progenitor in the future.

Some runaways could be traced back to the center of Barnard's loop, with a similar travel time of $\sim$6 Myr. Gaia DR2 2894122519382314624, 3172086171347714048, and 3260593833727106304 are low mass pre-main sequence stars with the ages of $\sim$6--9 Myr (McBride et al. in prep). They are located $\sim$20--30$^\circ$ away, traveling with the speeds of 17--40 \kms\ relative to the Complex. One B star, HD 32320, can also be traced back to be near the epicenter over a comparable time frame. It is difficult to conclusively state, however, if the ejection of these sources is necessarily due to the supernova and not due to the regular cluster dynamics.

\software{TOPCAT \citep{topcat}, Plotly \citep{plotly}}

\acknowledgments
I thank Kevin Covey and Keivan Stassun on their comments regarding the manuscript. Furthermore, I thank Josefa Gro{\ss}schedl on the discussion regarding feedback in the Orion Complex, and Paul Plucinsky on the discussion regarding Monogem. M.K. acknowledges support provided by the NSF through grant AST-1449476, and from the Research Corporation via a Time Domain Astrophysics Scialog award (\#24217), as well as NASA ADAP grant 80NSSC19K0591.
This work has made use of data from the European Space Agency (ESA)
mission {\it Gaia} (\url{https://www.cosmos.esa.int/gaia}), processed by
the {\it Gaia} Data Processing and Analysis Consortium (DPAC,
\url{https://www.cosmos.esa.int/web/gaia/dpac/consortium}). Funding
for the DPAC has been provided by national institutions, in particular
the institutions participating in the {\it Gaia} Multilateral Agreement.

\bibliographystyle{aasjournal.bst}
\bibliography{ms.bbl}

\end{document}